# Shared-End-First Ancilla-Chain Ordering for Nested Multi-Controlled Cascades: A Reversible Binary-Addition-Tree Study

Wei-Chang Yeh

Department of Industrial Engineering and Engineering Management, National Tsing Hua University, Hsinchu 300044, Taiwan

Corresponding author: Wei-Chang Yeh (e-mail: yeh@ieee.org).

This work was supported in part by the National Science and Technology Council (NSTC), Taiwan, R.O.C., under Grants NSTC 114-2221-E-007-123-MY3 and NSTC 113-2221-E-007-117-MY3. The NVIDIA RTX PRO 6000 Blackwell Max-Q Workstation Edition GPU used for coherent-state validation was provided as in-kind hardware support through the NVIDIA Academic Grant Program.

**ABSTRACT** Control order is logically irrelevant for a multi-controlled-X gate but can become physically consequential after ordered ancilla-chain synthesis. This work derives and validates a shared-end-first ordering rule for nested-control cascades, using reversible Binary-Addition-Tree (BAT) successor circuits as the primary case. Under Maslov clean-v-chain synthesis, pre-optimization CX cost is order-invariant, whereas shared-end-first ordering exposes P(m)=(m-4)(m-3)/2 inverse relative-phase-CCX pairs across stage boundaries; standard optimization removes 6P(m) CX, giving 12m-31 for QBAT after an independently derived two-CX terminal identity. At m=20, retaining an ascending control list after mirroring the BAT direction yields 1,025 CX, while mirroring the chain orientation restores 209. A disjoint-target nested cascade reproduces the same cancellation count for count, showing that BAT target geometry is not required; an ordered dirty-v-chain control shows zero orientation advantage, delimiting the mechanism within the tested synthesis families. Independent ripple-carry baselines give 11m-28 CX on FULL connectivity, so the QBAT/RC ratio approaches 12/11, with equal natural width 2m-3 in the tested QBAT-A1 and INC_RC implementations. The contribution is a synthesis-ordering rule that exposes cancellation to an existing optimizer, not a new optimizer pass or a universal MCX rule.

# I. INTRODUCTION

Nested multi-controlled quantum circuits occur in arithmetic, comparator, oracle, and reversible state-generation constructions. Although an MCX operator is invariant under permutation of its control qubits, an ancilla-based decomposition can impose an ordered partial-product chain. That ordering can change the cross-gate structure exposed to a compiler even when the logical operator is unchanged. This article studies that representation effect in a reversible Binary-Addition-Tree (BAT) cascade, where consecutive MCX control sets are strongly nested and therefore provide a clean setting for separating synthesis cost from cross-stage optimization.

The BAT family provides structured state generation through simple coordinate-update rules. It originated in exact binary-state network reliability [1] and was later extended to multiterminal and parallel settings [3], [5], with related Quick BAT and QB-II variants [2], [4]. The present work uses BAT as a circuit-structured case study rather than as a claim of arithmetic novelty.

Two directional BAT transition conventions are central to the present study. In the **leftmost-zero rule**, the leftmost zero is changed to one and all coordinates with smaller indices are reset to zero. For three coordinates,

$$000 \rightarrow 100 \rightarrow 010 \rightarrow 110 \rightarrow 001 \rightarrow 101 \rightarrow 011 \rightarrow 111.$$

In the **rightmost-zero rule**, the rightmost zero is changed to one and all coordinates with larger indices are reset to zero:

$$000 \rightarrow 001 \rightarrow 010 \rightarrow 011 \rightarrow 100 \rightarrow 101 \rightarrow 110 \rightarrow 111.$$

The two rules enumerate the same $2^m$ binary states under opposite coordinate conventions. Their definitions are zero-selection and reset rules. Under appropriate positional labels, each transition is also equivalent to modular increment, but that arithmetic representation is a derived property rather than the operational BAT definition.

Quantum computation requires reversible evolution described by unitary operators [6], [7]. Reversible logic and multi-controlled gates provide a direct route from a bijective classical state transition to a quantum circuit [8], [13]–[16]. The corrected dual-rule BAT formulation is especially convenient because terminal wraparound from the all-one state to the all-zero state makes both directional BAT transitions cyclic permutations of the complete basis. They therefore admit direct in-place unitary realizations on the $m$ data qubits.

A logical MCX description alone does not determine physical cost. MCX gates admit decompositions with very different ancilla, CX, depth, and non-Clifford tradeoffs [20]–[25]. Recent constructions further expand this design space through conditionally clean ancillas, polylogarithmic-depth decompositions, and optimized T-count or T-depth formulations [33]–[36]. The present paper therefore makes a deliberately scoped claim: control-list orientation matters for the tested ordered clean-ancilla v-chain because it changes the inter-stage structure presented to an existing optimizer; the result is not assumed to hold for every MCX synthesis family.

Two exact BAT realizations are characterized. A0 uses ancilla-free MCX synthesis. A1 uses a clean-ancilla v-chain with a reusable ordered partial-product chain. A 4,860-row production study over m=4,...,30 separates pre-optimization synthesis cost from post-optimization cancellation across FULL, LINE, and HEAVY_HEX coupling graphs using Qiskit 2.5.2 [27] and SABRE [28]. Both BAT directions and both v-chain control-list orientations are tested, with A0 serving as a negative control. A separate 120-run FULL-connectivity generality study then crosses two nested-control circuit families, two ordered v-chain synthesis families, two orientations, five dimensions, and three seeds.

Because the BAT successor is modular increment under the positional encodings b_L and b_R, the study also compares the BAT realization with independent exact ripple-carry and optimized relative-phase carry incrementers, together with exact and approximate QFT baselines. This comparison is intentionally adversarial: QBAT is not proposed as a better general-purpose incrementer. A separate placement-policy sensitivity study distinguishes circuit cost from initial-placement and routing effects on constrained topologies.

The contributions are threefold. First, the paper formalizes the leftmost-zero and rightmost-zero BAT rules as cyclic reversible maps and proves the mirror relation U_L = B U_R B. Second, it derives and audits a shared-end-first ordering rule for nested-control cascades synthesized with a reusable clean-ancilla v-chain. The rule exposes P(m)=(m-4)(m-3)/2 cross-stage inverse relative-phase-CCX pairs and yields a closed-form favorable FULL count of 12m-31 for QBAT. A disjoint-target nested cascade reproduces the same 6P(m) CX reduction, while an ordered dirty-v-chain control produces zero orientation advantage, locating the mechanism in clean ordered partial-product reuse rather than BAT target geometry or ordered synthesis in general. Third, the paper benchmarks the construction against independent arithmetic incrementers and quantifies placement sensitivity, including negative results where specialized arithmetic remains cheaper.

The arithmetic equivalence also defines the paper's boundary. BAT successor semantics can be implemented by an incrementer plus a wire relabeling, so QBAT-A1 is not claimed to dominate specialized arithmetic. The engineering contribution is instead the verified BAT construction and the demonstration that caller-supplied control order can expose or suppress large inter-stage cancellation in one important clean-ancilla synthesis. Repeated successor application still gives no quantum search speedup; any algorithmic advantage would have to arise from later composition with problem structure [11], [12], [26].

# II. BACKGROUND AND RELATED WORK

## A. BAT Lineage and Dual Zero-Selection Rules

Let

$$X = (x_1, x_2, \dots, x_m), \qquad x_i \in \{0,1\},$$

with coordinates written from left to right in increasing index order. The state space is

$$\Omega_m = \{0,1\}^m, \qquad |\Omega_m| = 2^m.$$

For a nonterminal vector define

$$j_L(X) = \min\{i : x_i = 0\}, \qquad j_R(X) = \max\{i : x_i = 0\}.$$

The leftmost-zero transition $T_L$ is

$$x'_{j_L} = 1, \qquad x'_i = 0 \ (i < j_L), \qquad x'_i = x_i \ (i > j_L),$$

whereas the rightmost-zero transition $T_R$ is

$$x'_{j_R} = 1, \qquad x'_i = 0 \ (i > j_R), \qquad x'_i = x_i \ (i < j_R).$$

Dual BAT transition conventions for m = 3

Leftmost-zero BAT 000 → 100 → 010 → 110 → 001 → 101 → 011 → 111

leftmost 0 -> 1; all smaller-index coordinates -> 0

Rightmost-zero BAT 000 → 001 → 010 → 011 → 100 → 101 → 110 → 111

rightmost 0 -> 1; all larger-index coordinates -> 0

**FIGURE 1. Dual BAT transition conventions for m = 3. The leftmost-zero rule resets smaller-index coordinates, whereas the rightmost-zero rule resets larger-index coordinates.**

For reversible realization, the terminal state is closed by

$$T_L(1^m) = T_R(1^m) = 0^m.$$

This converts each finite BAT enumeration path into a cycle.

### B. Reversible Quantum Circuits and MCX Synthesis

A basis-state permutation defines a unitary permutation matrix [6], [7]. The Pauli-$X$ gate performs

$$X|0\rangle = |1\rangle, \qquad X|1\rangle = |0\rangle,$$

and a C^kX gate flips one target when all k controls are one. High-control MCX gates are not native to typical quantum processors, so practical implementations decompose them into one- and two-qubit gates, sometimes using clean, dirty, borrowed, or conditionally clean ancillas [8], [20]–[25], [33], [34]. Recent work also targets low depth and optimized non-Clifford cost for multi-controlled gates [35], [36]. These alternatives reinforce that MCX resource conclusions are synthesis-family dependent.

Two synthesis regimes define the primary BAT study. A0 uses an exact ancilla-free MCX synthesis. A1 uses the exact Maslov-style clean-ancilla v-chain selected by Qiskit 2.5.2, which builds an ordered chain of partial AND products. Although the logical MCX operator is symmetric in its controls, this decomposition preserves an ordered control list. Different caller-supplied orders can therefore expose different cross-stage inverse subsequences when nested MCX gates are composed. A later generality control repeats the ordering experiment on a disjoint-target nested cascade and on Qiskit's ordered dirty-ancilla v-chain. The present study does not claim a new optimizer pass or a universal MCX ordering rule; it measures when representation exposes structure that an existing optimization pipeline can remove. Circuit simplification itself belongs to the broader literature on template matching, peephole optimization, and algebraic or diagrammatic reduction [30]–[32].

### C. Positional Equivalence and Mirror Symmetry

For the rightmost-zero rule define

$$b_R(X) = \sum_{i=1}^{m} 2^{m-i}\, x_i,$$

and for the leftmost-zero rule define

$$b_L(X) = \sum_{i=1}^{m} 2^{i-1}\, x_i.$$

Then

$$b_R\big(T_R(X)\big) \equiv b_R(X) + 1 \ (\mathrm{mod}\ 2^m)$$

and

$$b_L\big(T_L(X)\big) \equiv b_L(X) + 1 \ (\mathrm{mod}\ 2^m).$$

Define the bit-reversal operator $B$ by

$$B|x_1 x_2 \cdots x_m\rangle = |x_m x_{m-1} \cdots x_1\rangle.$$

Reversal converts the leftmost-zero rule into the rightmost-zero rule and vice versa:

$$T_L = B \circ T_R \circ B.$$

The corresponding quantum operators therefore satisfy

$$U_L = B U_R B.$$

This logical mirror equivalence does not, by itself, determine the resources of every lower-level synthesis. If an ordered ancillary workspace is introduced, its orientation must be mirrored consistently with the data-register transformation to preserve the same optimization opportunities.

### D. Compilation and Connectivity

Circuit synthesis and routing were implemented in Qiskit 2.5.2 [27]. SABRE was used for layout and routing [28]. Three coupling classes were studied: fully connected (FULL), bidirectional linear nearest-neighbor (LINE), and connected subgraphs of a single distance-7 heavy-hex parent graph. Heavy-hex connectivity is representative of low-degree superconducting layouts and is supported directly by the Qiskit coupling-map framework [29].

For fair A0-A1 comparison, both modes were compiled on the same physical graph of width equal to the measured A1 width. A0 therefore retained access to unused physical sites as layout and routing space. This intentionally avoids giving A1 an artificial connectivity advantage from being compiled on a larger graph. Because later controls showed that initial placement can be as important as routing on constrained topologies, the main free-SABRE campaign was supplemented by deterministic structure-aware placement tests based on each circuit's pre-transpilation interaction structure.

## III. MATERIALS AND METHODS

### A. Logical QBAT Operators

Define

$$U_L|X\rangle = |T_L(X)\rangle, \qquad U_R|X\rangle = |T_R(X)\rangle.$$

Equivalently,

$$U_d = \sum_{X \in \Omega_m} |T_d(X)\rangle\langle X|, \qquad d \in \{L, R\}.$$

The leftmost-zero circuit updates targets from high index to low index:

**Algorithm 1. Leftmost-zero QBAT successor $U_L$.**
**Input:** data qubits $q_1, \dots, q_m$ storing $x_1, \dots, x_m$.
**Step 1.** For $k = m, m-1, \dots, 2$, apply $C^{k-1}X(q_1, \dots, q_{k-1}; q_k)$.
**Step 2.** Apply $X(q_1)$.
**Step 3.** Return the data register.

The rightmost-zero circuit is the coordinate mirror:

**Algorithm 2. Rightmost-zero QBAT successor $U_R$.**
**Input:** data qubits $q_1, \dots, q_m$ storing $x_1, \dots, x_m$.
**Step 1.** For $k = 1, 2, \dots, m-1$, apply $C^{m-k}X(q_{k+1}, \dots, q_m; q_k)$.
**Step 2.** Apply $X(q_m)$.
**Step 3.** Return the data register.

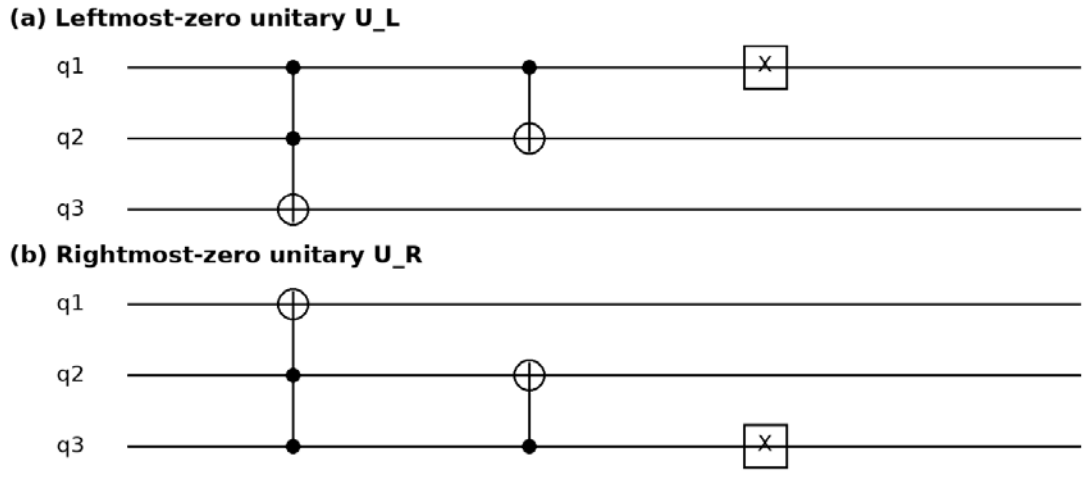

**FIGURE 2. Mirrored three-qubit QBAT-Basic circuits. U_L uses controls on smaller-index coordinates; U_R uses controls on larger-index coordinates.**

### B. A0 Ancilla-Free and A1 Clean-Ancilla Realizations

A0 uses the exact Qiskit MCXSynthesisNoAuxV24 synthesis (synth_mcx_noaux_v24) with zero ancillas. A1 uses synth_mcx_n_clean_m15, the non-deprecated exact Maslov-style clean-ancilla v-chain realization corresponding to MCXVChain(dirty_ancillas=False) [23]. The A1 implementation was independently checked for operator equivalence to the intended v-chain construction.

For an $n$-control MCX, the A1 synthesis requires $n-2$ clean ancillas. The largest QBAT gate has $m-1$ controls, so the maximum reusable pool is

$$a(m) = m - 3.$$

Because each A1 stage restores its clean ancillas, one pool can be reused across the complete cascade. The total algorithmic width is therefore

$$q_{A1} = m + a(m) = 2m - 3,$$

whereas

$$q_{A0} = m.$$

Clean-in/clean-out behavior was verified after explicit decomposition rather than assumed from the high-level synthesis contract.

### C. Control-Chain Orientation

For A1, the ordered list of logically symmetric controls is an experimental factor. ASC lists controls by increasing data-qubit index and DESC lists them by decreasing index. These labels describe caller-supplied order to the selected synthesis routine, not a claim about a universal Qiskit default. The target position and ancilla-pool order are held fixed, so the experiment isolates the control-chain orientation relative to the end shared by consecutive nested control sets.

- **ASC:** controls listed by increasing data-qubit index;
- **DESC:** controls listed by decreasing data-qubit index.

This yields four configurations: L-ASC, L-DESC, R-ASC, and R-DESC. The ancilla pool itself is supplied in a fixed ascending order. A fixed nominal control-order convention is not mirror-covariant: switching BAT direction while retaining ASC changes the ancillary-chain orientation relative to the shared end. The production study therefore treats direction and control-list order as separate factors rather than assuming that identical ASC/DESC labels represent mirrored implementations.

The nested leftmost-zero cascade has a common low-index end, while the rightmost-zero cascade has a common high-index end. The structural hypothesis tested in the production campaign was that a favorable v-chain orientation places the large shared control subset at the head of the ordered partial-product chain. Under this convention the expected mirrored pairs are

$$\text{L-ASC} \leftrightarrow \text{R-DESC}$$

and

$$\text{L-DESC} \leftrightarrow \text{R-ASC}.$$

No intrinsic cost preference was assigned to leftmost-zero or rightmost-zero BAT.

### D. Compilation Protocol

The production environment used Python 3.12.13 and Qiskit 2.5.2 on an Intel Core Ultra 7 265K workstation with 187 GiB RAM. The primary basis was

$$\{rz, sx, x, cx\},$$

with SABRE layout and SABRE routing, optimization level 3, no approximation, and qubits_initially_zero=False. Ten fixed transpiler seeds were used: 11, 29, 47, 71, 101, 131, 173, 211, 257, and 307.

The primary sweep covered

$$m = 4, 5, \dots, 30.$$

A0 included both BAT directions. A1 included both directions and both control orders. Across FULL, LINE, and HEAVY_HEX, the production dataset contains 4,860 configuration rows. Each row records a pre-optimization and post-optimization transpilation, giving 9,720 transpilation executions with zero failures and zero timeouts.

For every $m$, A0 and A1 were compiled onto the same physical target width

$$q_{target} = q_{A1} = 2m - 3.$$

The heavy-hex parent graph was fixed at distance $d = 7$. The primary connected subgraph was generated by deterministic breadth-first selection from a preregistered minimum-eccentricity root. Fixed-$q$ sensitivity was separately tested at $m = 12, 20, 30$ using four structurally selected roots.

### E. Separating Synthesis Cost from Optimizer Cancellation

For every A1 circuit, resource counts were measured both before optimization and after optimization. On FULL connectivity, where routing is absent, this isolates synthesis structure cleanly. Define the CX cancellation fraction

$$C_{CX} = \frac{N_{CX}^{(0)} - N_{CX}^{(3)}}{N_{CX}^{(0)}},$$

where $N_{CX}^{(0)}$ and $N_{CX}^{(3)}$ are the level-0 and level-3 CX counts. An analogous depth-reduction fraction was calculated.

A mechanism audit at $m \in \{8, 12, 16, 20\}$ also stored the decomposed circuits and counted adjacent mutually inverse gate subsequences to a fixed point. This enabled direct testing of whether repeated compute/uncompute structures cross cascade-stage boundaries.

### F. Correctness and GPU Validation

Correctness was tested at several levels. Exhaustive computational-basis verification covered $m = 4, \dots, 10$ for both BAT directions and both A1 control orders. For $m = 11, \dots, 30$, 1,000 reproducible random basis inputs per dimension were propagated through explicitly decomposed reversible circuits. The decomposition exposed $X$, CX, CCX, and relative-phase CCX operations so that data output and ancilla restoration were measured rather than assumed.

A mirror-consistency acceptance test was run before production. All 120 cases satisfied exact zero CX and depth differences with pre-routing DAG equality, confirming the intended mirrored circuit and layout conventions.

For compiled-output validation, all 720 sweep rows within the tractable physical-width range were individually checked after undoing the recorded layout and routing permutation. Larger sweep rows were explicitly labeled in the machine-readable dataset as not individually verified because of scale rather than being implicitly counted as verified.

Large coherent-state validation used NVIDIA cuQuantum cuStateVec on an RTX PRO 6000 Blackwell Max-Q Workstation Edition. The tested widths were $q = 27,29,31$, selected only after a memory gate requiring at least 1.10 times a two-statevector working estimate. Logical output ordering was restored using transpiler layout metadata before comparison with the independent BAT amplitude permutation.

### G. Independent Arithmetic Baselines and Predicate-Attachment Control

Because $b_L$ and $b_R$ map the BAT successor to modular increment, an independent baseline study implemented exact ripple-carry increment (INC_RC), an optimized exact relative-phase carry construction (INC_OPT), exact QFT increment (INC_QFT_EXACT), and three preregistered approximate-QFT variants, drawing on established quantum-arithmetic constructions [9], [10], [17]–[19], [24], [25]. All exact methods implemented

$$|x\rangle \mapsto |x + 1 \bmod 2^m\rangle$$

and were verified independently of the QBAT cascade. The hardware-oriented comparison used the same Qiskit 2.5.2 basis, optimization level, seeds, and frozen coupling maps as the production QBAT study. COMMON-width and natural-width controls were also evaluated where the arithmetic architecture had a smaller natural register. The arithmetic campaign completed 11,880 successful configuration rows, corresponding to 23,760 level-0/level-3 transpilation executions, with zero failed runs and zero correctness failures.

A targeted composition test then attached the same prefix predicate $p_k$ to QBAT-A1, INC_RC, and INC_OPT. The composed transformation was

$$|x\rangle|0_{work}\rangle|z\rangle \mapsto |x + 1 \bmod 2^m\rangle|0_{work}\rangle|z \oplus p_k(x)\rangle.$$

This test was designed to distinguish merely exposing an intermediate Boolean value from being able to use that value coherently while restoring workspace and phase.

### H. Placement-Policy Sensitivity

The primary production campaign used free SABRE layout and routing. A targeted sensitivity study compared this policy with deterministic structure-aware placement on LINE and HEAVY_HEX. The structure-aware policy was derived from the pre-transpilation two-qubit interaction graph rather than selected from post-transpilation CX counts. It was applied to QBAT-A1, the arithmetic baselines, and finally A0 as a control. The A0 control used the same physical widths and coupling maps as the original A0-A1 comparison. This design tests whether an apparent sparse-topology penalty is architectural, routing-related, placement-related, or a mixture of these effects.

### I. Generality and Synthesis-Family Controls

To distinguish BAT geometry from nested-control structure, a FULL-connectivity generality study used m in {8,12,16,20,24} and seeds {11,47,101}. The reference family was QBAT-A1. The second family retained the same nested control sets but moved every stage target to a disjoint target register, so no target was a control of any stage. This breaks BAT target geometry while preserving the common-end nesting that motivates the ordering rule.

Each family was synthesized with two exact ordered-chain routines exposed by Qiskit 2.5.2: the clean-ancilla synth_mcx_n_clean_m15 realization used by A1 and the dirty-ancilla synth_mcx_n_dirty_i15 realization [37]. For both routines, reversing the caller-supplied control list reverses the induced ordered chain, so orientation is a meaningful experimental factor. Pre-optimization circuits were serialized before either optimization path was applied. A conditionally-clean Khattar-Gidney family was inspected but not used as evidence because its fixed one- or two-ancilla construction does not expose a length-(n-2) caller-indexed partial-product chain.

The resulting matrix contains 120 successful runs with zero seed variance in every reported count. Clean-v-chain correctness was checked exhaustively for m=8 and 12 and by reproducible random propagation for larger m; dirty-v-chain primitives were verified against MCX tensor identity for arbitrary dirty-ancilla states, with end-to-end statevector checks at m=8. Stage-boundary audits at m=12 and 20 iterated inverse-pair removal to a fixed point.

## IV. CORRECTNESS AND RESOURCE ANALYSIS

### A. Unitarity and Correctness

**Proposition 1.** $U_L$ and $U_R$ are unitary.

*Proof.* Under $b_L$ and $b_R$, respectively, each BAT transition increments one label modulo $2^m$. Hence both $T_L$ and $T_R$ are bijections of $\Omega_m$. Their matrix representations are permutation matrices, so

$$U_L^\dagger U_L = U_R^\dagger U_R = I. \quad ▫$$

**Proposition 2.** Algorithm 1 implements $T_L$ exactly and Algorithm 2 implements $T_R$ exactly.

*Proof.* For $T_L$, let $j = j_L(X)$. All coordinates $x_i$ with $i < j$ equal one. Gates targeting $k > j$ are blocked by the original $x_j = 0$. The gate targeting $j$ has only lower-index controls, all one, and therefore activates $x_j$. Subsequent lower-index targets are flipped from one to zero. The all-one case activates every stage and completes the wraparound. The proof for $T_R$ is the mirrored argument. ▫

### B. Mirror Theorem and Operational Corollary

**Theorem 1.** Let $B$ reverse the data-qubit order. Then

$$T_L = B \circ T_R \circ B$$

and

$$U_L = B U_R B.$$

*Proof.* Bit reversal maps the leftmost zero to the rightmost zero and maps the smaller-index reset set to the corresponding larger-index reset set. The all-zero and all-one states are invariant under reversal, so the terminal closure is preserved. Linearity gives the quantum relation. ▫

**Corollary 1.** The two logical BAT circuits have identical logical gate count, maximum control width, serial logical depth, and full-cycle length under mirrored construction. For the clean-ancilla v-chain realization, the same resource symmetry is recovered when the ordered ancillary chain is mirrored consistently with the data-register transformation.

The second sentence is operational rather than purely algebraic. A fixed ascending control convention applied to both directions does not represent the same ancillary-chain orientation relative to the cascade. The resource comparison must therefore use mirrored chain orientation, not only mirrored data labels.

### C. Full-Cycle Property

**Theorem 2.** Repeated application of either $U_L$ or $U_R$ from $|0^m\rangle$ visits every computational-basis state exactly once and returns to $|0^m\rangle$ after $2^m$ applications.

*Proof.* Under the corresponding positional label, each transition increments by one modulo $2^m$. The $2^m$ labels are distinct before wraparound. ▫

### D. Logical and Algorithmic Resources

At the logical MCX level, each BAT transition contains one $C^kX$ block for every $k = 0,1,\dots,m-1$. Thus the total logical gate count is $m$, the maximum control width is $m-1$, and the serial logical depth is $m$.

**TABLE I**
**LOGICAL AND ALGORITHMIC RESOURCES OF A0 AND A1.**

| Quantity | A0 ancilla-free | A1 clean-ancilla |
|---|---|---|
| Data qubits | m | m |
| Clean ancillas | 0 | m - 3 |
| Algorithmic width | m | 2m - 3 |
| Logical MCX blocks | m | m |
| Maximum controls | m - 1 | m - 1 |
| Control-order factor | not primary | ASC or DESC |
| Ancilla pool reuse | n/a | one shared clean pool |

For the resource tradeoff, define

$$G_{CX} = \frac{N_{CX}^{A0} - N_{CX}^{A1}}{N_{CX}^{A0}}, \qquad G_D = \frac{D^{A0} - D^{A1}}{D^{A0}},$$

and per-added-ancilla efficiencies

$$E_{CX} = \frac{N_{CX}^{A0} - N_{CX}^{A1}}{m-3}, \qquad E_D = \frac{D^{A0} - D^{A1}}{m-3}.$$

As a secondary engineering indicator, the algorithmic-width-depth product is

$$WD = q_{alg}\,D.$$

This quantity is not presented as a universal quantum-computing cost function.

### E. Closed-Form Resource Accounting for Shared-End-First A1

Under the selected Maslov-style clean-ancilla decomposition, a k-control MCX block with k >= 2 contains one exact CCX and 2(k-2) relative-phase CCX operations. In the {rz, sx, x, cx} basis this gives 6k-6 CX gates; the k=1 stage is one CX. Summing the BAT cascade therefore gives the pre-optimization A1 count N_CX,pre(m) = 1 + sum_{k=2}^{m-1}(6k-6) = 3m² - 9m + 7. This is a decomposition identity, not a regression fit.

For shared-end-first ordering, the structural audit finds 1,2,...,m-4 adjacent inverse relative-phase-CCX pairs across successive stage boundaries, for P(m)=(m-4)(m-3)/2 pairs. Each inverse pair removes six CX gates after basis decomposition. A separate m-independent terminal identity removes two more CX gates at the k=2/k=1/k=0 junction: the tail of the exact CCX, the k=1 CX, and the terminal X form a two-qubit block on data qubits 0 and 1 whose Weyl coordinates are exactly (pi/4,0,0), so its three CX are locally equivalent to one CX plus one-qubit rotations. Equivalently, CNOT conjugation maps the target-Z rotation to a ZZ rotation and then back to a single-qubit Z rotation under the crossed CNOT. This identity was localized at the same structural position for m=8,9,12 and is valid for all m>=4. Hence N_CX,fav(m)=3m^2-9m+7-6P(m)-2=12m-31. The shared-end-last orientation has no audited cross-stage inverse pairs and remains N_CX,unfav(m)=3m^2-9m+5 after the same terminal identity. All FULL-connectivity seeds agree with these counts at every tested m.

A model-dependent non-Clifford count follows from the same structure. Using a conventional seven-T exact CCX and four-T relative-phase CCX model [21]–[23], the pre-optimization A1 cascade has 4m² - 13m + 10 T gates and shared-end-first cancellation reduces this to 15m - 38. These are decomposition-model counts, not a complete fault-tolerant cost comparison: T-depth, magic-state scheduling, and a common Clifford+T accounting for INC_RC and INC_OPT were not part of the frozen baseline dataset.

## V. RESULTS

### A. Correctness and Dual BAT Equivalence

The production verification reported zero correctness failures. Exhaustive verification covered $m = 4,\dots,10$, random decomposed-circuit verification covered $m = 11,\dots,30$, and coherent tests covered representative widths. Every tested A1 stage restored its shared clean-ancilla pool to $|0\cdots0\rangle$.

The mirror-consistency acceptance test passed 120/120 cases exactly. During this preproduction gate, a non-mirror-covariant A0 control-order convention was detected and corrected before any production data were collected. The full sweep contained 720 individually verified transpiled outputs and 4,140 larger outputs explicitly marked as not individually verified because of scale; no unverified row was silently counted as verified. The measured A1 width was exactly

$$q_{A1} = 2m - 3$$

for all 27 production dimensions.

### B. Ancilla-Free and Clean-Ancilla Resource Scaling

Figure 3 compares A0 with the favorable A1 orientation across the three connectivity classes under the preregistered free-SABRE protocol. The A1 advantage grows rapidly after the smallest dimensions.

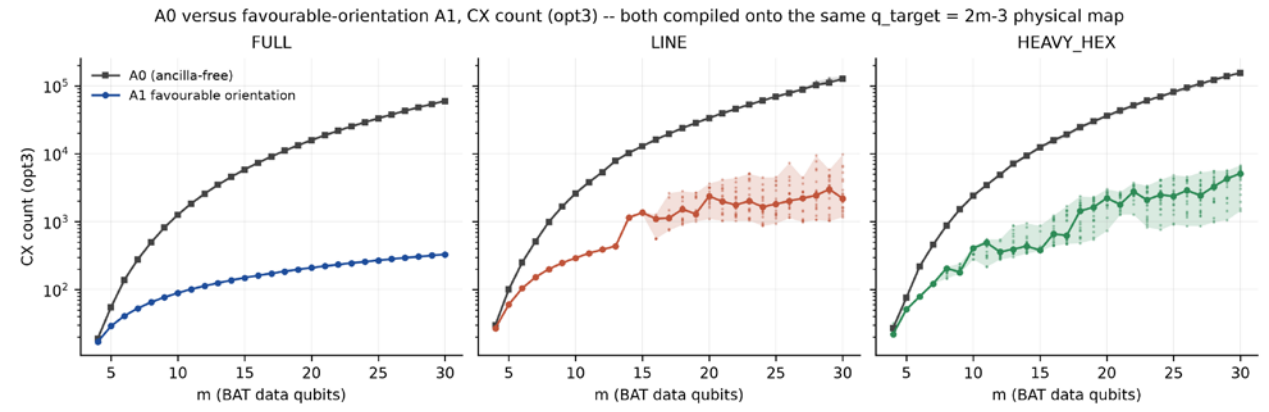

**FIGURE 3. Median CX count versus BAT dimension for A0 and favorable A1 realizations under FULL, LINE, and HEAVY_HEX connectivity using the primary free-SABRE compilation policy.**

Across the complete free-SABRE sweep, favorable A1 produced topology-level median CX gains of 98.1% on FULL, 93.1% on LINE, and 94.6% on HEAVY_HEX relative to A0. Median depth gains were 98.1%, 95.3%, and 94.8%, respectively. These percentages remain useful as an A0 ablation, but they are not used as the principal external benchmark because A0 is an intentionally ancilla-free high-control synthesis.

On FULL connectivity, the compiled CX and depth counts are seed-invariant at each fixed m. The FULL medians quoted above summarize variation across the tested dimensions rather than seed-to-seed compiler variability.

A placement-policy sensitivity control later applied the same structure-aware treatment to A0. A0 did not benefit: its median CX changed by -17.3% on LINE and -6.2% on HEAVY_HEX, meaning the constrained placement made A0 more expensive. By contrast, A1 improved by 59.7% and 9.8%, respectively. On the matched $m = 12,16,20,24,30$ subset, the A1-over-A0 CX gain changed from 94.5% to 98.6% on LINE and from 95.3% to 96.4% on HEAVY_HEX. Thus the original free-SABRE A0 comparison was not artificially favorable to A1.

The algorithmic-width-depth product is retained only as a secondary engineering indicator in the reproducibility package and is not used as a headline cost metric.

### C. Control-Chain Orientation and Inter-Stage Cancellation

The strongest mechanistic result appears on FULL connectivity, where routing cannot contaminate the pre-optimization comparison. All four A1 direction/order configurations have exactly the same pre-optimization CX cost. The closed-form accounting in Section IV-E explains this common cost and predicts the favorable 12m-31 count from the number of stage-boundary inverse relative-phase-CCX pairs. The resource difference therefore arises from cross-stage optimization opportunity rather than cheaper synthesis of any individual MCX block.

$m = 20$, all four configurations start at 1,027 CX gates. After optimization, L-ASC and R-DESC fall to 209 CX gates, whereas L-DESC and R-ASC remain at 1,025. Thus naively reversing BAT direction while retaining a fixed ascending control list changes the optimized count from 209 to 1,025; reversing the chain orientation together with the data convention restores 209. The favorable orientation removes 79.6% of the pre-optimization CX count, while the unfavorable orientation removes only 0.2%.

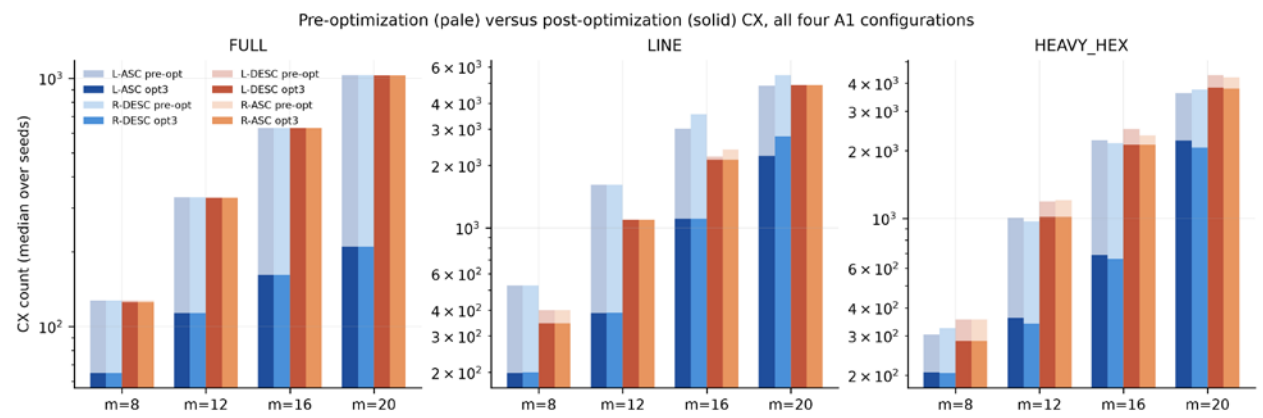

**FIGURE 4. Pre-optimization and post-optimization CX counts for selected A1 configurations. The four configurations have equal synthesis cost on FULL, but only the shared-end-first orientation exposes extensive inter-stage cancellation.**

The optimization-level ladder shows that the cross-stage favorable cancellation is already fully realized at optimization level 1. The terminal two-CX identity follows two compiler pathways: in the favorable orientation the relevant CNOT pair becomes directly cancellable at level 1, whereas in the unfavorable orientation the equivalent 3-CX-to-1-CX block requires two-qubit resynthesis and therefore appears only at optimization level 2 or higher. Levels 2 and 3 provide no additional CX reduction for the favorable circuits in the mechanism subset.

**TABLE II**
**OPTIMIZATION-LEVEL MECHANISM AUDIT ON FULL CONNECTIVITY.**

| Case | CX level 0 | CX level 1 | CX level 2 | CX level 3 |
|---|---|---|---|---|
| m = 8, L-ASC | 127 | 65 | 65 | 65 |
| m = 8, L-DESC | 127 | 127 | 125 | 125 |
| m = 12, L-ASC | 331 | 113 | 113 | 113 |
| m = 12, L-DESC | 331 | 331 | 329 | 329 |
| m = 16, L-ASC | 631 | 161 | 161 | 161 |
| m = 16, L-DESC | 631 | 631 | 629 | 629 |
| m = 20, L-ASC | 1,027 | 209 | 209 | 209 |
| m = 20, L-DESC | 1,027 | 1,027 | 1,025 | 1,025 |

The structural audit detected adjacent inverse subsequences crossing cascade-stage boundaries. At $m = 20$, the favorable orientations contained 136 such inverse pairs, all crossing a stage transition; the unfavorable orientations contained zero. Across the mechanism audit, all 520 detected inverse pairs crossed stage boundaries.

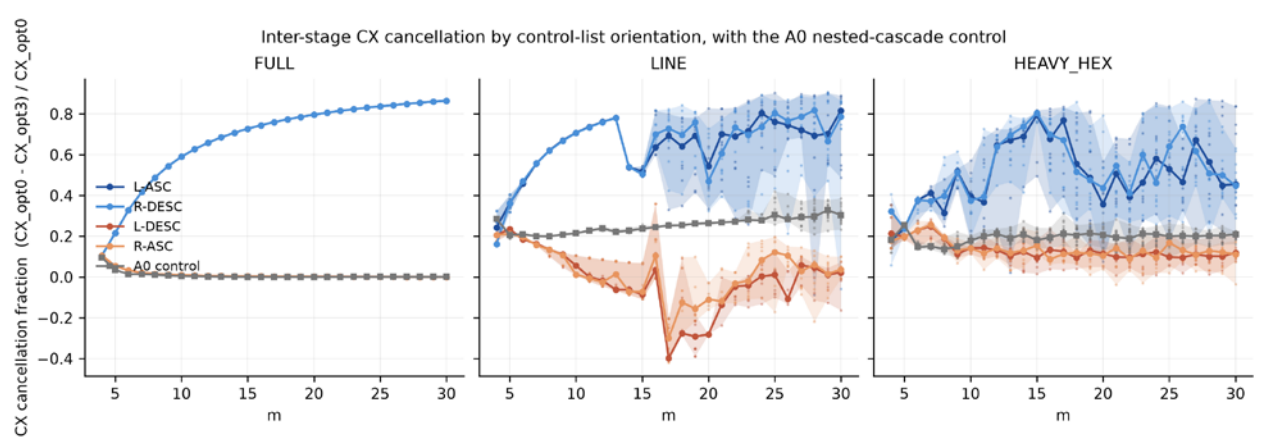

**FIGURE 5. CX cancellation fraction versus m for the four A1 control-chain orientations and the A0 control. On FULL, favorable A1 cancellation reaches 86.5% at m = 30, while the A0 mechanism control remains at or below 1.2% in the audited subset.**

The A0 negative control is important. A0 has the same nested high-control cascade structure but no ordered ancillary partial-product chain. On FULL it exhibits a median opt0-to-opt3 CX cancellation of approximately 0.1%, with at most 1.2% in the mechanism subset. The large cancellation is therefore associated with alignment of the ordered ancillary chain rather than nesting alone. This observation is related to, but distinct from, general circuit simplification and pattern-matching methods [30]–[32]: the contribution here is the circuit-ordering rule that exposes the cancellable structure, not a new cancellation pass.

### D. Generality Across Cascade Geometry and Synthesis Family

The generality study separates target geometry from clean-chain reuse. The disjoint-target family uses exactly the same nested control sets as QBAT, but every stage target lies in a separate register and is never a control of any stage. Under clean ordered v-chain synthesis, this non-BAT family reproduces the ordering effect count for count. At m=20, both orientations begin at 1,026 CX; shared-end-first falls to 210, whereas shared-end-last remains at 1,026. The 816-CX advantage is identical to the cross-stage component of QBAT at the same m.

Closed-form accounting confirms the match. For the separate-target family, N_CX,pre=3m^2-9m+6, N_CX,first=12m-30, and N_CX,last=N_CX,pre. At m=12 and 20, the fixed-point audit finds exactly P(m)=36 and 136 cross-stage inverse relative-phase-CCX pairs, respectively, with zero within-stage pairs and exactly 6P(m) removable CX. Target placement is therefore not required for the mechanism; the common-end nesting and reusable clean partial-product chain are sufficient in the tested families.

The dirty ordered v-chain [37] gives the complementary negative control. Caller order is a real synthesis variable and pre-optimization counts remain orientation-invariant, but both orientations have identical post-optimization cost in both circuit families and expose zero cross-stage inverse pairs. For example, at m=20 the dirty QBAT realization is 1,397 CX before optimization and 1,395 after in both orientations, while the separate-target realization remains 1,396 CX. The count result is deterministic across the tested seeds; full end-to-end dirty-circuit statevector validation was limited to m=8, with larger cases supported by exact primitive verification and structural audits. Thus the observed ordering advantage is not a generic property of ordered MCX synthesis within the tested families; it depends on a clean reusable partial-product chain that recomputes the same intermediate products across adjacent nested stages.

TABLE III
GENERALITY CONTROL FOR SHARED-END-FIRST ORDERING ON FULL CONNECTIVITY.

| Family / synthesis | Pre-order invariant? | m=20 CX first / last | Cross-stage pairs |
|---|---|---|---|
| QBAT-A1 / clean | Yes | 209 / 1,025 | 136 |
| Separate-target / clean | Yes | 210 / 1,026 | 136 |
| QBAT-A1 / dirty | Yes | 1,395 / 1,395 | 0 |
| Separate-target / dirty | Yes | 1,396 / 1,396 | 0 |

The conditionally-clean kg24 routines were not used as positive or negative evidence because they do not expose the required caller-indexed partial-product chain. The resulting claim is therefore deliberately bounded: shared-end-first ordering generalizes beyond BAT target geometry under the tested clean ordered v-chain, but it does not generalize to the tested dirty ordered v-chain.

### E. Comparison with Independent Quantum Incrementers

The arithmetic baseline study directly addresses the fact that, under b_L or b_R, a BAT successor is increment by one up to wire relabeling. Exact ripple-carry (INC_RC), optimized exact relative-phase carry (INC_OPT), exact QFT, and approximate-QFT baselines were implemented independently under the same toolchain. INC_OPT was the lowest-cost exact arithmetic baseline across the measured topologies; INC_RC provides the most transparent carry-based comparison. Thus the BAT construction is evaluated as a structured reversible realization, not as a claim of arithmetic superiority.

On FULL connectivity, every transpiler seed gives the same CX count at a fixed m. For favorable QBAT-A1, Section IV-E derives N_CX = 12m - 31 from the selected v-chain decomposition and the audited stage-boundary cancellations. The independently implemented INC_RC baseline gives the exact observed relation N_CX = 11m - 28 throughout m=4,...,30. The ratio is therefore

$$N_{CX}^{QBAT-A1} = 12m - 31,$$
$$N_{CX}^{INC_RC} = 11m - 28.$$

The ratio implied by these measured relations is

$$R_{CX}(m) = \frac{12m - 31}{11m - 28},$$

which approaches 12/11 = 1.0909. At m=30, favorable QBAT-A1 requires 329 CX gates versus 302 for INC_RC, a ratio of 1.089. Both measured constructions use 57 algorithmic qubits at this dimension, while the QBAT/INC_RC depth ratio is 1.433. Thus FULL-connectivity CX cost is within about 9% of ripple carry, but QBAT is not the least-cost incrementer and near parity does not extend to depth.

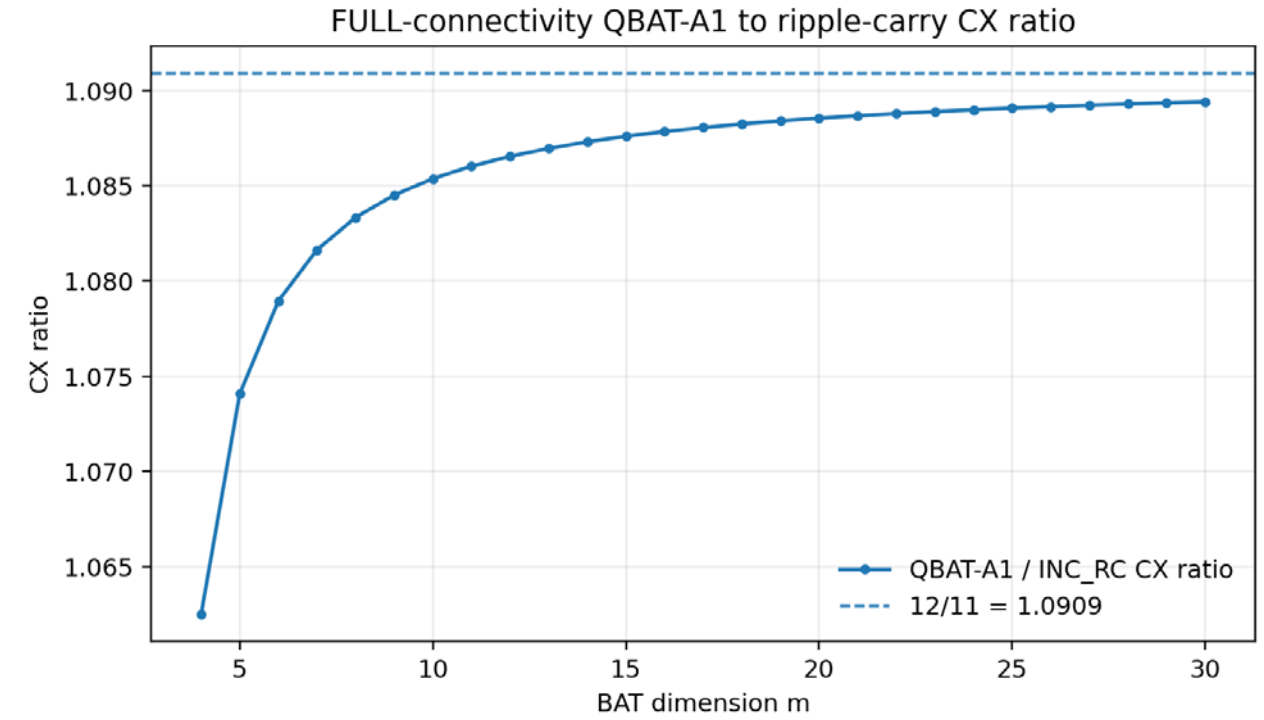

**FIGURE 6. FULL-connectivity CX ratio of favorable QBAT-A1 to INC_RC for m = 4,...,30. The m = 30 value is part of a flat approach to the ratio implied by the measured linear relations rather than an isolated endpoint.**

TABLE IV
SELECTED m = 30 ARITHMETIC COMPARISONS UNDER THE PRIMARY FREE-LAYOUT POLICY.

| Topology | QBAT CX | RC CX | CX ratio | Depth ratio | Widths Q/R/target |
|---|---|---|---|---|---|
| FULL | 329 | 302 | 1.089 | 1.433 | 57 / 57 / 57 |
| LINE | 2,196.5 | 1,002 | 2.149 | 2.133 | 57 / 57 / 57 |
| HEAVY_HEX | 5,101.5 | 927.5 | 5.654 | 6.607 | 57 / 57 / 57 |

QBAT-A1 and INC_RC both use m data qubits plus m-3 clean ancillas in the tested implementations, so their natural algorithmic widths coincide exactly at 2m-3; at m=30 both use 57 qubits. The common 2m-3-site target is also a compilation control used for matched connectivity. This equality is implementation-specific and is not a lower bound for quantum incrementers: lower-ancilla increment constructions exist [10], [24] but were not instantiated in the present baseline set. The width coincidence also does not extend to all tested arithmetic baselines; INC_QFT_EXACT has natural width m and occupies the common target with idle sites.

Table IV reports seed medians. Half-integer CX entries arise only because the median of ten integer-valued transpilation outcomes can lie midway between the two central observations; they do not represent fractional gates.

The targeted placement study is reported separately because it used a different initial-layout policy. Under structure-aware placement, the measured $m = 30$ QBAT/INC_RC CX ratios decreased to 1.665 on LINE and 2.132 on HEAVY_HEX.

A separate predicate-attachment test found no measured structural-cost advantage for QBAT. For the composed increment-plus-prefix-predicate operation at $m = 30$, $k = 28$, on FULL, the measured CX counts were 195 for INC_OPT, 303 for INC_RC, and 330 for QBAT. The marginal predicate operation added one CX and one flag qubit to each architecture and required no extra Toffoli. INC_OPT used its relative-phase carry directly and passed coherent validation without a repair stage. The predicate test therefore preserves the bare arithmetic ordering rather than creating a QBAT advantage.

### F. Placement Policy, Routing, and Sparse Connectivity

The large free-SABRE gaps on sparse graphs were not purely architectural. A structure-aware placement sensitivity study substantially reduced the QBAT cost while leaving the arithmetic baselines much less affected. At $m = 30$, QBAT-A1 on LINE fell from 2,196 to 869 CX gates, and its ratio to INC_RC decreased from 2.149 to 1.665. On HEAVY_HEX,

QBAT-A1 fell from 5,102 to 1,610 CX gates and the ratio decreased from 5.654 to 2.132.

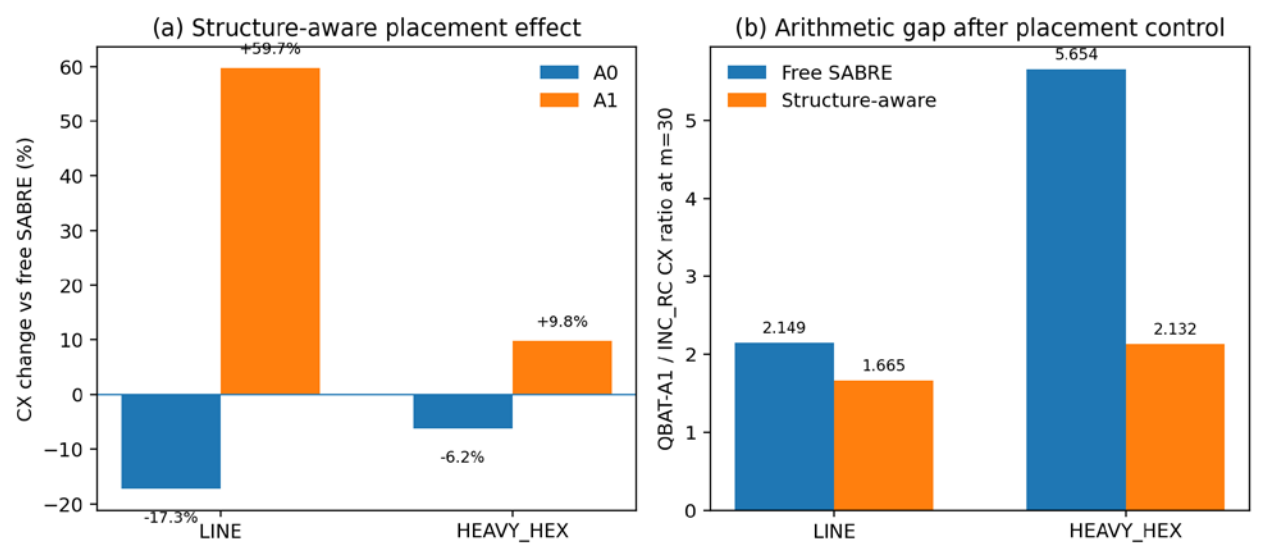

**FIGURE 7. Placement-policy sensitivity. (a) Structure-aware placement improves A1 but worsens A0 under the tested policy, reflecting their different interaction structures. (b) At m = 30, structure-aware placement substantially reduces the apparent QBAT-to-ripple-carry CX gap on LINE and HEAVY_HEX.**

The A0 control clarifies why structure-aware placement is not universally beneficial. A0's nested high-control MCX gates generate a dense interaction structure with little one-dimensional locality to exploit, so constraining the initial placement removes useful SABRE freedom. A1 contains a genuine ordered ancillary chain for which contiguous or interaction-aware placement is exploitable. Under the matched sensitivity control, A0 CX became 17.3% worse on LINE and 6.2% worse on HEAVY_HEX, whereas A1 improved by 59.7% and 9.8%.

Accordingly, sparse-topology cost is best interpreted as a mixture of architecture, initial placement, and subsequent routing. The original 2.149 and 5.654 ratios should not be called intrinsic topology penalties.

### G. Heavy-Hex Subgraph Variability

The heavy-hex study used one fixed distance-7 parent graph and a preregistered primary root. To test sensitivity to subgraph choice, the same fixed-$q$ experiments at $m = 12,20,30$ were repeated over four structurally selected roots.

The median relative CX spread across roots was 0.1649, compared with a median seed spread of 0.1149, and the maximum root spread reached 0.8914 for one configuration. Root choice can therefore move individual heavy-hex results substantially.

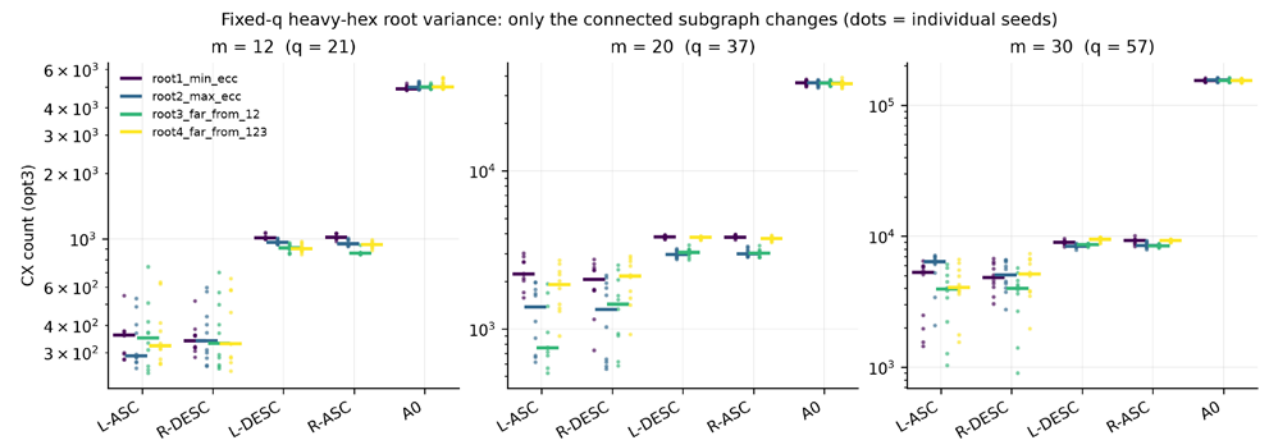

**FIGURE 8. Fixed-q heavy-hex root/subgraph variability at representative dimensions.**

The coarse large-scale trend in favorable A1 gain from small to large $m$ exceeds root variation and is robust. Fine point-to-point heavy-hex structure at $m \geq 10$, however, is comparable to variation caused by changing the connected subgraph. Local fluctuations are therefore not interpreted as a heavy-hex scaling law.

### H. GPU-Assisted Coherent-State Validation and Reproducibility

Seven large A1 circuits were validated with cuQuantum cuStateVec on the RTX PRO 6000. The tested widths included $q = 27$, 29, and 31. All launched cases passed with fidelity equal to one to displayed precision, phase-aligned $L_2$ error on the order of $10^{-15}$, and ancilla-zero probability equal to one.

**TABLE V**
**SELECTED GPU COHERENT-STATE VALIDATION RESULTS.**

| m | Config. | q | CX | Sim. time | Phase-aligned L2 |
|---|---|---|---|---|---|
| 15 | L-ASC | 27 | 149 | 1.2 s | 1.96e-15 |
| 15 | L-DESC | 27 | 545 | 3.5 s | 4.30e-15 |
| 16 | L-ASC | 29 | 161 | 5.1 s | 1.91e-15 |
| 17 | L-ASC | 31 | 173 | 21.9 s | 2.03e-15 |

At $m = 15$, favorable and unfavorable orientations implemented the same target unitary despite 149 versus 545 CX gates. The predicate-attachment follow-up also performed coherent checks, including the relative-phase carry baseline, with no phase-restoration failures.

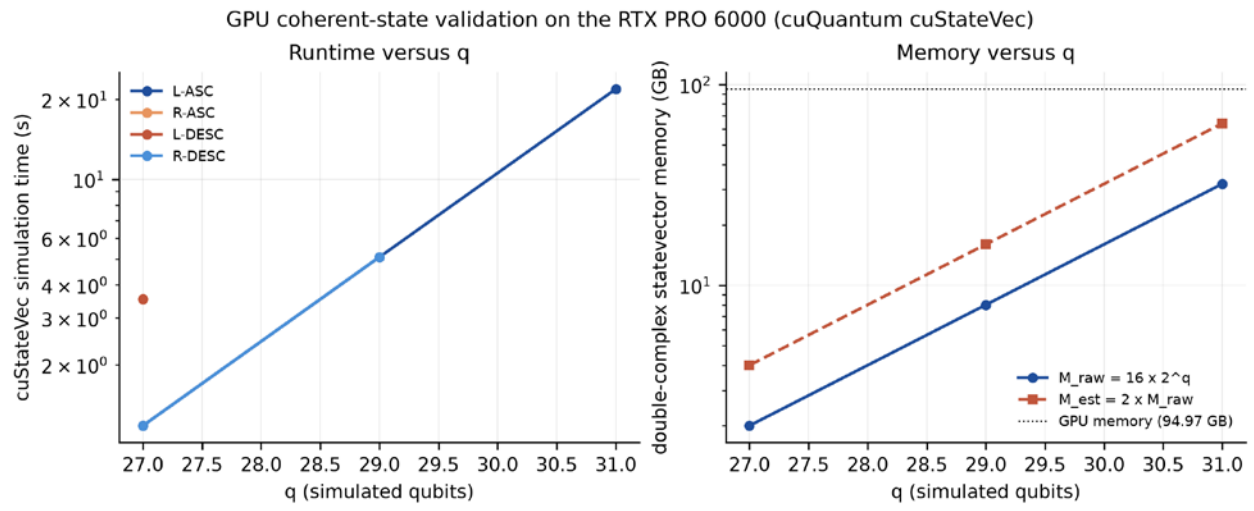

**FIGURE 9. GPU simulation runtime and statevector memory for the validated A1 cases.**

The main production package contains source code, frozen coupling maps, raw CSV files, QPY circuits, environment records, tables, figures, logs, and a SHA-256 manifest. The arithmetic-baseline and targeted placement/predicate studies were maintained as separate reproducibility packages so that the original production data remained frozen.

## VI. DISCUSSION

### A. What the Study Establishes

The study establishes results at three levels. At the logical level, the two canonical BAT successors are cyclic permutation operators related by bit reversal. At the synthesis level, the selected clean-ancilla v-chain turns a logically irrelevant control permutation into a physically relevant representation choice across a nested cascade: shared-end-first ordering exposes inverse stage-boundary structure that an existing optimizer can remove. At the generality-control level, the same cancellation reproduces count for count when the stage targets are moved to a disjoint register, but disappears under an ordered dirty-ancilla v-chain. The mechanism is therefore not BAT-target-geometry-specific and not universal across ordered MCX synthesis.

The favorable mirrored implementation must therefore mirror the orientation of the ancillary partial-product chain as well as the data-register labels. This conclusion is supported by exact FULL mirror checks, four-way control-order experiments, the analytical CX accounting, and the A0 negative control.

### B. Inter-Stage Cancellation in Context

The mechanism is not that one v-chain orientation is intrinsically cheaper to synthesize. Before optimization, all four direction/order configurations have identical FULL CX counts. Shared-end-first ordering aligns repeated partial products across consecutive nested stages, exposing inverse compute/uncompute subsequences. Qiskit's existing optimizer removes those subsequences, and the measured CX benefit is already complete at optimization level 1.

Accordingly, this work is a representation and synthesis-ordering result rather than a proposal for a new optimizer pass. General quantum-circuit optimization already includes template matching, peephole simplification, continuous-parameter methods, and algebraic or diagrammatic reduction [30]–[32]. The generality control identifies the conditions supported by the present evidence: consecutive stages must have nested control sets sharing a common end, synthesis must create an ordered clean-ancilla partial-product chain reused across stages, and the shared end must be placed at the head of that chain. Target geometry is not among the required conditions: the disjoint-target cascade gives the same P(m) pair count and 6P(m) CX reduction. Conversely, the tested dirty ordered v-chain gives zero orientation advantage because each stage restores arbitrary incoming dirty-ancilla contents within the stage, leaving no reusable partial product exposed at the boundary. Recent conditionally-clean constructions [33] use different internal structures and are not claimed to support or refute this rule.

Relation to explicit partial-product hoisting. The repeated partial products identified by the stage-boundary audit could alternatively be retained deliberately across adjacent stages in a custom cascade-level construction, omitting their uncompute/recompute pairs by design. The audit gives exact gate accounting for that alternative: retaining the shared clean partial products removes the same 6P(m) CX term exposed by shared-end-first ordering. At m=20, P(m)=136, so explicit hoisting of the audited pairs removes 816 CX from the 1,027-CX decomposed cascade, leaving 211 CX before the independent terminal two-CX identity. The present contribution addresses a different engineering route: ordinary stage-by-stage MCX synthesis is kept unchanged, and control ordering exposes the same cross-stage reuse to a stock optimizer, which reaches 209 CX after the terminal identity as well. Thus partial-product hoisting itself is not claimed as new, and no claim is made that the stock optimizer is superior to a custom hoisted circuit; the result is that the hoistable structure can be recovered without a bespoke cascade synthesizer.

The A0 control strengthens this interpretation from another direction. A0 has the same nested logical BAT cascade but lacks the ordered clean partial-product chain; on FULL it exhibits only negligible cancellation. Together, A0 and the dirty-v-chain control show that nesting and caller-visible ordering alone are insufficient. The large effect requires the clean reusable partial-product structure audited in A1 and the separate-target family.

### C. Quantified Relation to Quantum Arithmetic

Under b_L and b_R, both BAT rules are modular increment, and b_L versus b_R is only a wire relabeling. Therefore an incrementer followed by the corresponding wire permutation is an equally valid implementation of the successor semantics. QBAT-A1 is not proposed as a superior arithmetic object. Its measured contribution is the BAT-structured realization and the ordering behavior of its ancillary chain; specialized arithmetic is cheaper in the tested baseline study.

The FULL result is nevertheless quantitatively close in CX. The favorable QBAT-A1 count is derived as 12m-31 and all ten FULL seeds agree at each tested m. INC_RC gives 11m-28 throughout the same range. Their ratio approaches 12/11, and at m=30 it is 329/302 = 1.089. This is a near-parity CX result under unrestricted connectivity, not arithmetic superiority; the depth ratio at the same point is 1.433, and INC_OPT remains cheaper overall.

The predicate-attachment experiment further limits the BAT-specific claim. INC_RC exposes the same prefix carry predicate at no material marginal cost, and INC_OPT can use its relative-phase carry directly without repair in the tested composed operation. QBAT therefore does not have a measured unique predicate-attachment advantage. Its value in the present paper is the BAT-coordinate-preserving construction and the resource behavior of its ordered ancillary realization, not superiority over specialized arithmetic.

### D. Initial Placement, Routing, and Connectivity

The constrained-topology experiments demonstrate that compiler policy must be separated into initial placement and routing. Under free SABRE, the $m = 30$ QBAT/INC_RC CX ratios were 2.149 on LINE and 5.654 on HEAVY_HEX. Structure-aware placement reduced them to 1.665 and 2.132, respectively. The original larger gaps therefore cannot be described as intrinsic architecture penalties.

The control is especially informative because placement affects A0 and A1 in opposite directions under the tested policy. A1 has an ordered chain with exploitable locality, whereas A0's dense high-control interaction graph offers little comparable locality. Constraining A0's initial placement worsened its CX count, while the same class of structure-aware treatment improved A1. Structure-aware placement is therefore not universally beneficial; it is useful when the circuit exposes a physical interaction structure worth preserving.

Heavy-hex subgraph choice remains another material factor. Fixed-$q$ root variance is comparable to compiler seed variation and can be much larger for individual configurations. Fine local heavy-hex fluctuations should not be elevated to scaling laws.

### E. Practical Implementation Guidance

Four implementation rules follow from the measurements. First, retain the explicit BAT coordinate convention and its regression tests. Second, when using the tested clean ordered v-chain on a nested-control cascade, place the controls shared by consecutive stages at the head of the chain; in the BAT indexing this is L-ASC and R-DESC. This is a concrete mirroring hazard: at m=20 on FULL, L-ASC and R-DESC each compile to 209 CX, whereas reversing the BAT direction but retaining ASC produces R-ASC at 1,025 CX. Third, compare against independent arithmetic baselines on matched physical targets. Fourth, treat initial placement as an explicit experimental variable on constrained topologies. The study does not claim that a universal Qiskit default is unfavorable; it identifies a caller-order convention that can fail to preserve the intended chain orientation.

The finding that optimization level 1 captures the favorable cancellation is practically useful for the tested Qiskit stack. It indicates that the main cancellation does not require aggressive level-3 resynthesis. The numerical result remains compiler-version-specific and should be rechecked when the synthesis or optimization pipeline changes.

### F. Limitations and Research Scope

QBAT-Basic remains a successor primitive. Sequential traversal of the complete BAT cycle still requires $2^m$ applications, so the present resource reductions do not constitute an algorithmic search speedup. Quantum advantage, if any, must arise from later composition with problem predicates and quantum search, minimum-finding, or other structured procedures [11], [12], [26].

The native resource results are conditional on Qiskit 2.5.2, the selected exact A0 and clean/dirty ordered v-chain syntheses, the {rz, sx, x, cx} basis, compiler seed, coupling graph, initial-layout policy, and routing policy. The FULL-only generality control establishes that the clean-chain cancellation is not specific to BAT target geometry, but it also shows that the effect disappears for the tested dirty ordered v-chain. End-to-end dirty-v-chain statevector verification was performed at m=8; the larger dirty-v-chain instances exceeded the chosen statevector cap, so their deterministic resource counts rely on separately verified MCX tensor I synthesis primitives and structural circuit analysis rather than full end-to-end statevector simulation. Recent MCX constructions use conditionally clean ancillas, borrowed ancillas, polylogarithmic-depth recursions, or alternative T-optimal structures [33]–[36]. In particular, the inspected kg24 routines do not expose the caller-indexed chain required by the present ordering rule, so they were treated as outside the mechanism test rather than as a negative control. The numerical magnitude of the effect should not be transferred to another synthesis family without a direct structural test.

The arithmetic study also clarifies scope. QBAT-A1 is close to ripple carry in FULL CX cost but remains more expensive than the tested specialized arithmetic circuits overall, particularly after accounting for depth and constrained connectivity. A predicate-attachment control also found no QBAT-specific marginal-cost advantage. The present paper therefore does not propose QBAT as a replacement for a general-purpose incrementer.

Finally, 720 of the 4,860 production sweep rows were individually verified at the physical-circuit level. Larger rows were not individually statevector-verified because the target width reached 57 qubits. Their logical constructions were nevertheless verified exhaustively or randomly through $m = 30$, and representative compiled A1 circuits were coherently validated up to 31 qubits on the RTX PRO 6000.

## VII. CONCLUSIONS

This paper uses reversible Binary-Addition-Tree successor circuits to derive and test a shared-end-first ordering rule for nested multi-controlled cascades synthesized with an ordered clean-ancilla chain. The two BAT directions are cyclic permutation operators related by exact bit reversal, providing a controlled primary case in which logically equivalent control orders can be compared without changing the target unitary.

$$U_L = BU_RB.$$

Two exact realizations were characterized: A0, an ancilla-free MCX synthesis, and A1, a clean-ancilla Maslov v-chain requiring a reusable pool of m-3 ancillas and total algorithmic width 2m-3. The favorable A1 orientation places the controls shared by consecutive nested stages at the head of the partial-product chain.

The central implementation result is shared-end-first inter-stage cancellation. All clean-v-chain orientations have the same pre-optimization FULL cost, N_CX,pre=3m^2-9m+7 for QBAT, but the favorable orientation exposes P(m)=(m-4)(m-3)/2 inverse relative-phase-CCX pairs across stage boundaries. Together with the independently derived two-CX terminal identity, this yields N_CX,fav=12m-31. A fixed ascending convention is not mirror-covariant: at m=20 the optimized count changes from 209 to 1,025 when the BAT direction is reversed without reversing the chain, and returns to 209 when the chain orientation is mirrored as well.

A separate disjoint-target nested cascade reproduces the clean-chain cancellation exactly: its favorable reduction is 6P(m) CX with the same P(m) pair count as QBAT. An ordered dirty-v-chain control exposes a genuine caller-ordering knob but yields zero cross-stage inverse pairs and zero orientation advantage in both circuit families. The supported principle is therefore broader than BAT target geometry but narrower than ordered MCX synthesis in general: it applies to the tested nested cascades when a reusable clean partial-product chain recomputes the same shared-end products across consecutive stages.

Independent arithmetic baselines place the result in context. Exact ripple carry gives N_CX = 11m-28 on FULL, so the favorable QBAT/ripple-carry ratio approaches 12/11 and is 329/302 = 1.089 at m=30. This is a strong unrestricted-connectivity CX result but not arithmetic superiority: QBAT has higher depth, INC_OPT is cheaper overall, and a predicate-attachment control shows no unique QBAT marginal-cost advantage.

Sparse-topology measurements are strongly compiler-policy-dependent. Structure-aware placement reduced the $m = 30$ QBAT/INC_RC CX ratio from 2.149 to 1.665 on LINE and from 5.654 to 2.132 on HEAVY_HEX. A matched A0 control showed that the same placement treatment can worsen a dense interaction structure while benefiting the ordered A1 chain. Initial placement and routing must therefore be reported separately when interpreting constrained-connectivity overhead.

Correctness was established through exhaustive basis tests, random decomposed-circuit verification, coherent tests, exact mirror-consistency checks, GPU statevector validation up to 31 qubits, and dedicated clean/dirty generality controls. The resulting contribution is therefore deliberately scoped: a verified BAT-structured reversible construction, a closed-form ordering mechanism generalized beyond BAT target geometry for the tested clean v-chain, a negative synthesis-family boundary under the dirty v-chain, and an external resource characterization against arithmetic and placement controls.

## DATA AND CODE AVAILABILITY

The complete reproducibility material is supplied with the submission. It includes the frozen production archive, arithmetic-baseline and placement/predicate controls, the A0 structure-aware control, the residual-two-CX derivation package, and the FULL-only generality study with serialized pre-optimization circuits. Source code, frozen coupling maps, raw transpilation and correctness data, circuit artifacts, GPU validation records, environment specifications, tables, figures, logs, and SHA-256 manifests are included. No external dataset is required.

## ACKNOWLEDGMENT

The NVIDIA RTX PRO 6000 Blackwell Max-Q Workstation Edition GPU used for coherent-state validation was provided as in-kind hardware support through the NVIDIA Academic Grant Program.